\documentclass{iau} 
\usepackage{graphicx}

\title[WN3/O3 Wolf-Rayet Stars] %% give here short title %%
{The Evolutionary Status of WN3/O3 Wolf-Rayet Stars}

\author[Neugent, Massey, Hillier and Morrell]   %% give here short author list %%
{Kathryn F. Neugent$^{1,2}$, Phil Massey$^{1,2}$, D. John Hillier$^3$\\ \and Nidia I. Morrell$^4$}

\affiliation{$^1$Lowell Observatory, 1400 W Mars Hill Road, Flagstaff, AZ 86001\\ email: {\tt kneugent@lowell.edu}, {\tt phil.massey@lowell.edu} \\[\affilskip]
$^2$Department of Physics and Astronomy, Northern Arizona University,\\ Flagstaff, AZ, 86011-6010\\[\affilskip] $^3$Department of Physics and Astronomy, University of Pittsburgh, Pittsburgh, PA 15260\\ email: {\tt hillier@pitt.edu}\\[\affilskip]$^4$ Las Campanas Observatory, Carnegie Observatories, Casilla 601, La Serena, Chile\\ email: {\tt nmorrell@lco.cl}}

\pubyear{2017}
\volume{329}
\jname{The lives and death-throes of massive stars}
\editors{J.J. Eldridge, ed.}
\begin{document}

\maketitle

% 5 pages total
\begin{abstract}
As part of a multi-year survey for Wolf-Rayet stars in the Magellanic Clouds, we have discovered a new type of Wolf-Rayet star with both strong emission and absorption. While one might initially classify these stars as WN3+O3V binaries based on their spectra, such a pairing is unlikely given their faint visual magnitudes. Spectral modeling suggests effective temperatures and bolometric luminosities similar to those of other early-type LMC WNs but with mass-loss rates that are three to five times lower than expected. They additionally retain a significant amount of hydrogen, with nitrogen at its CNO-equilibrium value (10$\times$ enhanced). Their evolutionary status remains an open question. Here we discuss why these stars did not evolve through quasi-homogeneous evolution. Instead we suggest that based on a link with long-duration gamma ray bursts, they may form in lower metallicity environments. A new survey in M33, which has a large metallicity gradient, is underway. 
\end{abstract}

\firstsection

\section{Introduction}
A few years ago we began a search for Wolf-Rayet (WR) stars in the Magellanic Clouds \cite[(Massey et al. 2014, 2015, 2016)]{MCWRs,MCWR15,MCWR16}. As I write this proceeding, we have just finished our forth year of observations and have imaged the entire optical disks of both the Large Magellanic Cloud (LMC) and the Small Magellanic Cloud (SMC). We will likely have a few candidate WRs left to spectroscopically confirm after this season of imaging, but so far our efforts have been quite successful. We have discovered four WN-type stars, one WO-type star, eleven Of-type stars, and ten stars that appear to belong to an entirely new class of WR star.

\begin{figure}[h!]
\begin{center}
\includegraphics[width=2.5in]{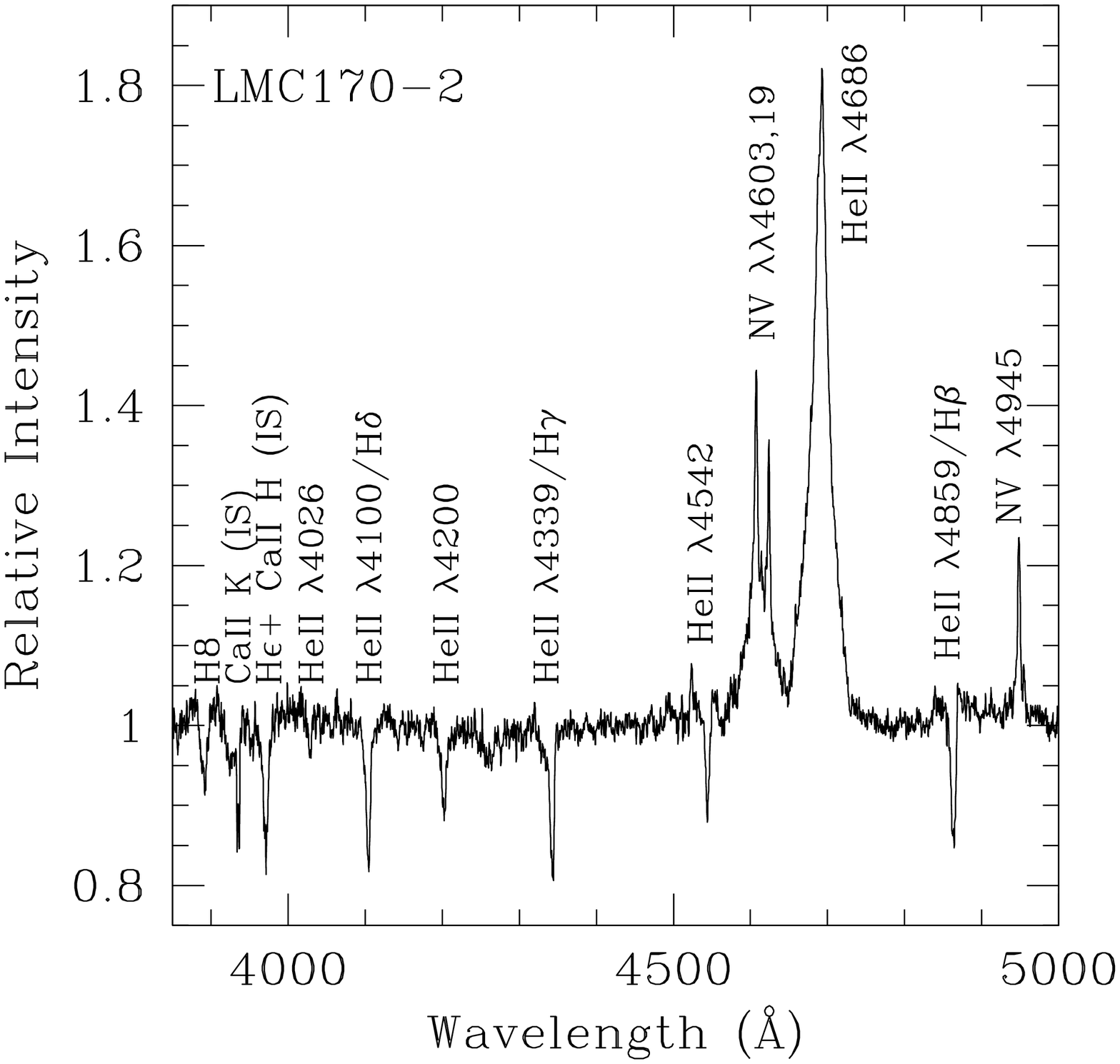} 
\includegraphics[width=2.5in]{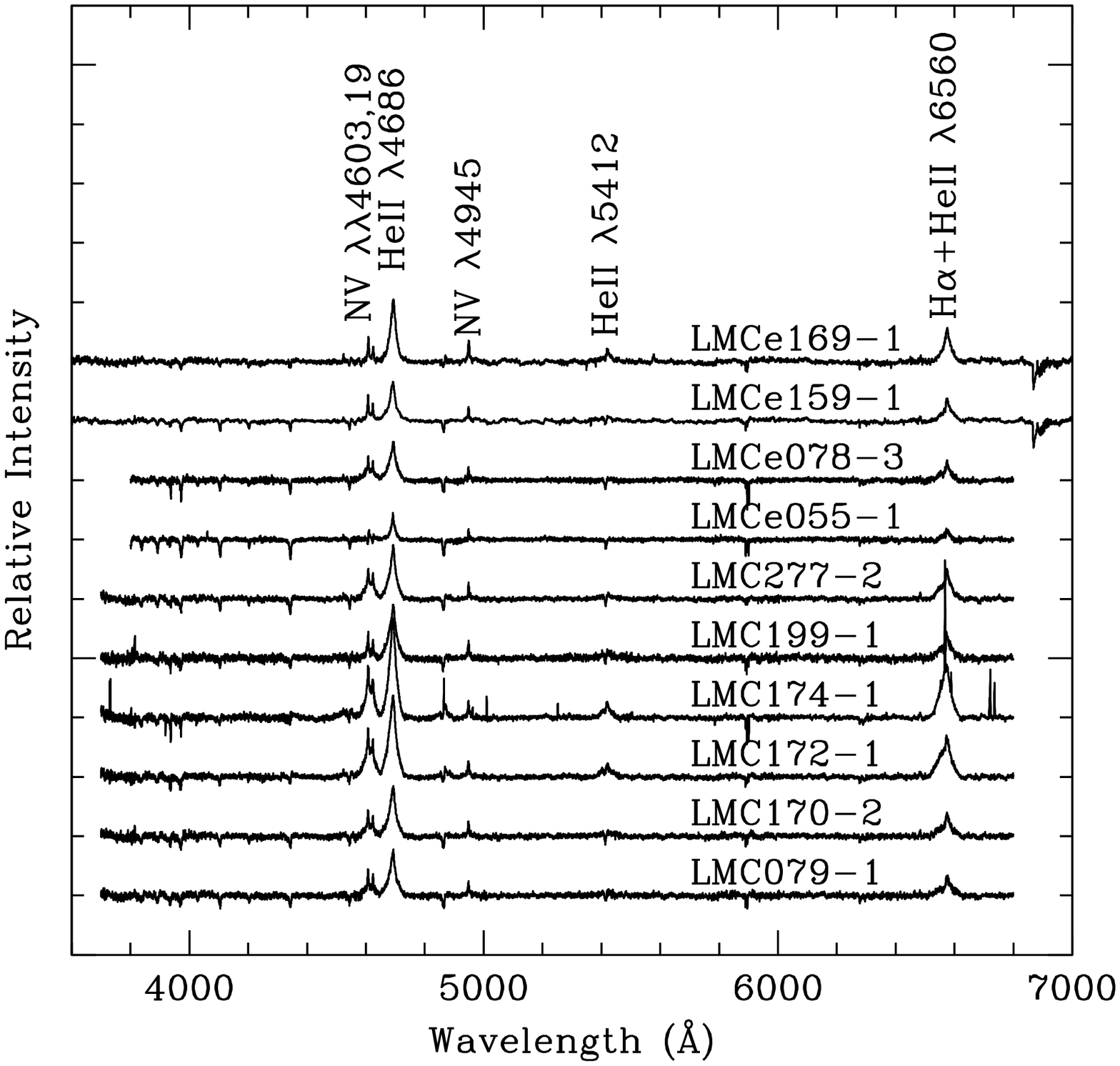} 
\caption{Left: The spectrum of LMC170-2, a WN3/O3 Wolf-Rayet star. Notice both the strong emission and absorption lines. Right: The spectra of all ten WN3/O3s.}
     \label{fig:spectrum}
\end{center}
\end{figure}

The ten stars, as shown in Figure~\ref{fig:spectrum} (left), exhibit strong WN3-like emission line features as well as O3V star absorption lines. While one might instinctively think these are WN3 stars with OV3 binary companions, this cannot be the case. They are faint with $M_V \sim -2.5$. An O3V by itself has an absolute magnitude of $M_V \sim -5.5$ \cite[(Conti 1988)]{Conti88} so these stars cannot contain an O-type star companion. We call these stars WN3/O3s where the ``slash" represents their composite spectra. As mentioned previously, we have found ten of these stars, or 6.5\% of the LMC WR population. As Figure~\ref{fig:spectrum} (right) shows, all of their spectra are nearly identical.

\section{Modeling Efforts and Physical Parameters}
Given that these stars are not WN3 stars with O3V companions, we were interested to see if we could model their spectra as single stars. We obtained medium dispersion optical spectra for all ten WN3/O3s in addition to high-dispersion optical spectra for one WN3/O3, UV spectra for three WN3/O3s, and NIR spectra for one WN3/O3. For LMC170-2 we obtained spectra from all four sources giving us coverage from 1000-2500\AA. To model these stars we turned to {\sc cmfgen}, a spectral modeling code that contains all of the complexities needed to model hot stars near their Eddington limits \cite[(Hillier \& Miller 1998)]{CMFGEN}. Using {\sc cmfgen}, we began by modeling the spectrum of LMC170-2.

\begin{figure}[h!]
\begin{center}
\includegraphics[width=2.2in]{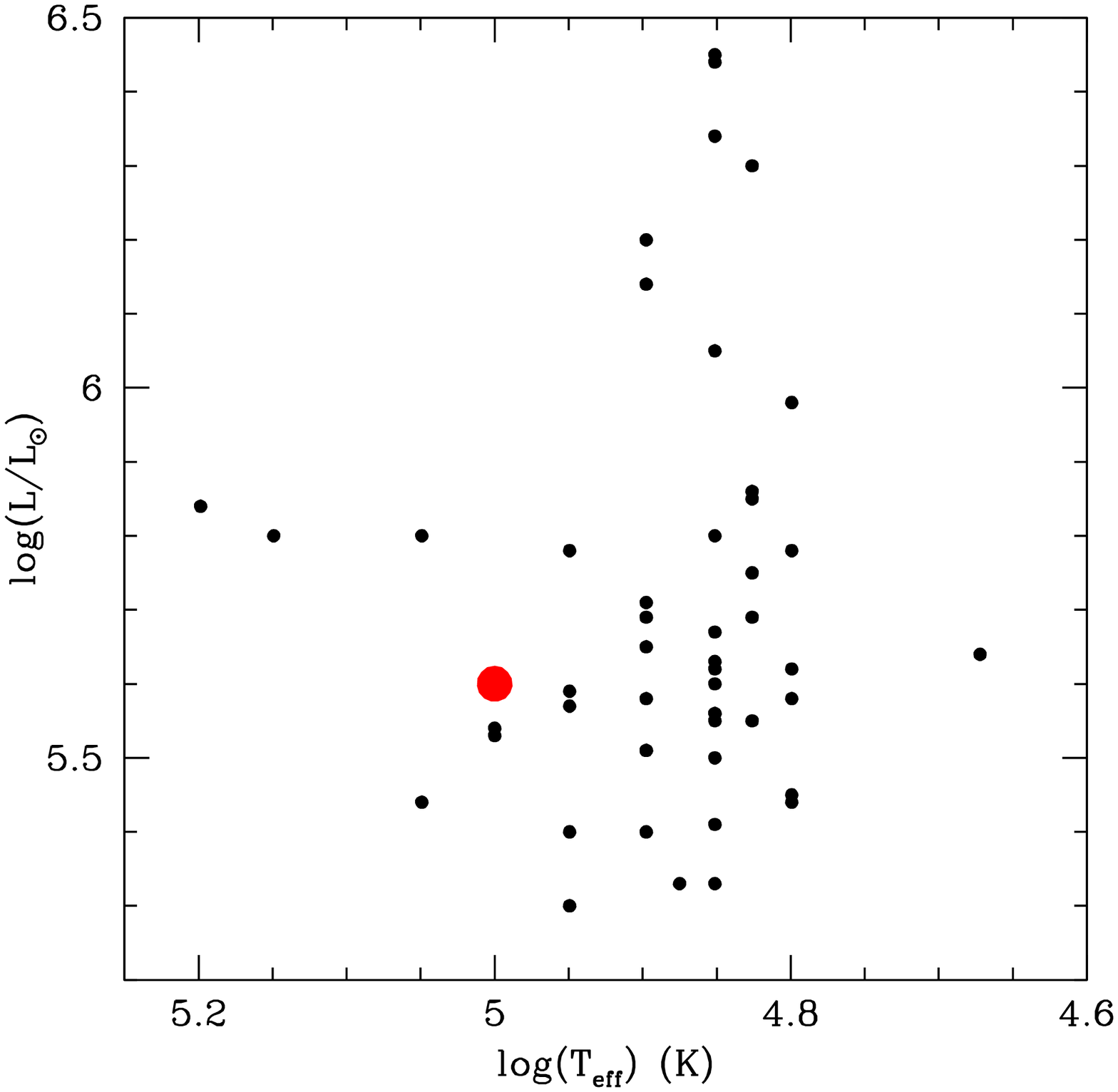} 
\includegraphics[width=2.2in]{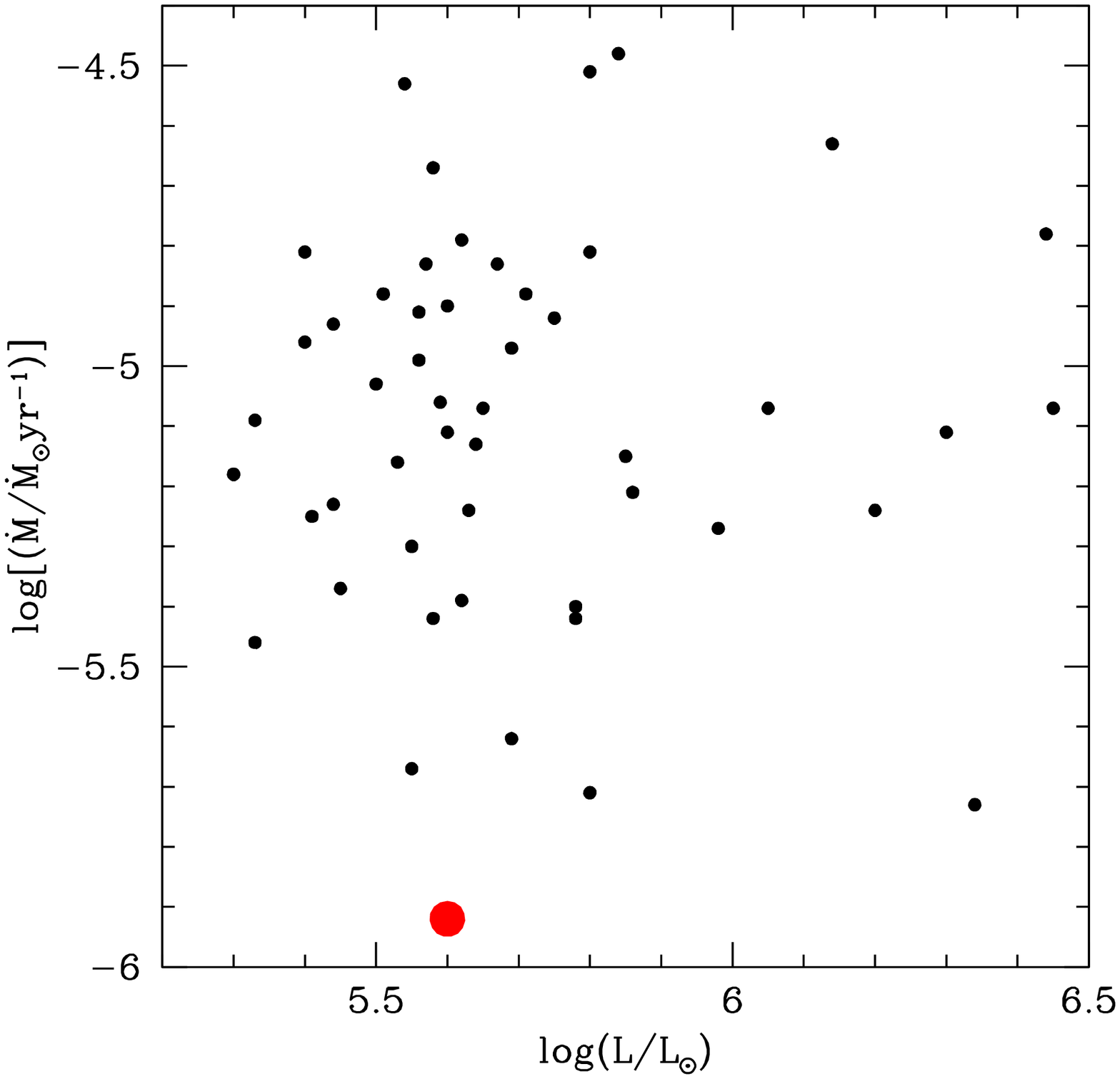} 
\caption{Left: A comparison of the bolometric luminosity vs.\ effective temperature for LMC170-2. Right: A comparison of the mass-loss rate vs.\ bolometric luminosity for LMC170-2. The small black circles represent the early-type LMC WNs (WN3s and WN4s) analyzed by \cite[Hainich et al. (2014)]{Potsdam}. LMC170-2 is represented as a large red circle.} 
     \label{fig:pots}
\end{center}
\end{figure}

Using the parameters given in Table~\ref{tab:params}, we obtained a good fit to both the emission and the absorption lines for LMC170-2. Hainich et al.\ (2014) recently modeled most of the previously known WN-type WRs in the LMC using PoWR and thus we can compare the star's physical parameters with those of more ``normal" WN-type LMC WRs. We can additionally compare the parameters with those of LMC O3Vs. Table~\ref{tab:params} shows that the abundances and He/H ratio are comparable with LMC WN3s. Figure~\ref{fig:pots} (left) shows that while the temperature is a bit on the high side of what one would expect for a LMC WN, it is still within the expected temperature range. However, the biggest surprise is the mass-loss rate. As is shown in Table~\ref{tab:params}, the mass loss rate of the WN3/O3s is more similar to that of an O3V than of a normal LMC WN. This is shown visually in Figure~\ref{fig:pots} (right). 

\begin{figure}[h!]
\begin{center}
\includegraphics[width=2.5in]{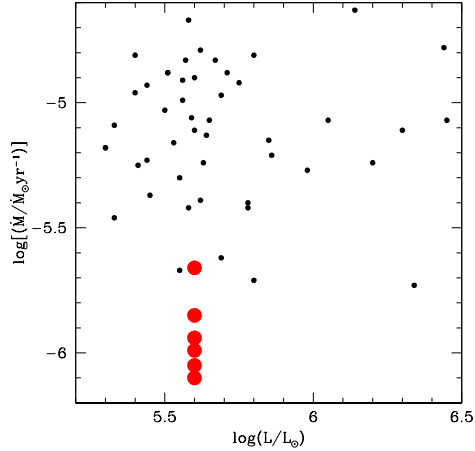} 
\caption{A comparison of the mass-loss rate vs.\ luminosity for all ten WN3/O3s. The small black circles represent the early-type LMC WNs (WN3s and WN4s) analyzed by \cite[Hainich et al. (2014)]{Potsdam}. The WN3/O3s are represented as large red circles.} 
     \label{fig:Mdot}
\end{center}
\end{figure}

\begin{table}[h!]
  \begin{center}
  \caption{Physical parameters of WN3/O3s, WNs, and O3Vs in the LMC$^a$}
  \label{tab:params}
  \begin{tabular}{l | l l l}
  \hline
& WN3/O3 & WN3 & O3V \\ \hline
$T_{\rm{eff}} (K)$ & 100,000 & 80,000 & 48,000\\
$\log \frac{L}{L_{\odot}}$ & 5.6 & 5.7 & 5.6\\
$\log \dot{M}^b$ & -5.9 & -4.5 & -5.9 \\
He/H (by \#) & 1.0 & 1.0-1.4 & 0.1\\
N (by mass) & 10$\times$ solar & 5-10$\times$ solar & 0.5$\times$ solar \\ 
M$_V$ & -2.5 & -4.5 & -5.5 \\\hline
  \end{tabular}\\
 \scriptsize{
 a: {\cite[Hainich et al. (2014)]{Potsdam}; \cite[Massey et al. (2013)]{Ostars}}\\
 b: {$\log M$ assumes a clumping filling factor of 0.1}}
  \end{center}
\end{table}
While all the spectra look visually similar, we still wanted to determine a range of physical parameters for these WN3/O3s. Thus, we modeled all 10 of them using {\sc cmfgen}. As discussed above, we obtained UV data for three of our stars. Based on these data, we determined that C\,{\sc iv} $\lambda$1550 was not present in the spectra of our stars. This again confirms that there is not an O3V star within the system. Additionally, it points to a high temperature as this line disappears at an effective temperature at 80,000 K. As an upper limit, as the effective temperature increases above 110,000 K, both He\,{\sc ii} $\lambda$4200 and O\,{\sc vi} $\lambda$1038 become too weak. Thus, the overall temperature regime is between 80,000 - 110,000 K. However, in practice, our models only varied between 100,000 - 105,000~K. Most of the other physical parameters all stayed relatively consistent and are thus well constrained as is shown in Table~\ref{tab:paramRange}. However, the exception is again the mass-loss rates. Figure~\ref{fig:Mdot} shows the range in the mass-loss rates for our WN3/O3s. While most of them are quite low (like those of O3Vs), a few of them are on the low end of some other early-type LMC WNs. In particular, there are three LMC WNs with low mass-loss rates and similar luminosities to the WN3/O3s. However, these three stars are visually much brighter than the WN3/O3s and thus we do not believe that they are the same type of star. We are still planning on studying them further. 

\begin{table}[h!]
  \begin{center}
  \caption{Physical parameter range of WN3/O3s}
  \label{tab:paramRange}
  \begin{tabular}{l | l}
  \hline
$T_{\rm{eff}} (K)$ & 100,000 - 105,000\\
$\log \frac{L}{L_{\odot}}$ & 5.6 \\
$\log \dot{M}$* & -6.1 - -5.7 \\
He/H (by \#) & 0.8 - 1.5\\
N (by mass) & 5-10$\times$ solar \\ \hline
  \end{tabular}\\
 \scriptsize{
 *stellar wind parameters: clumping filling factor of 0.1, terminal velocity of 2,600 km s$^{-1}$, and $\beta$ = 1.}
  \end{center}
\end{table}

\section{Evolutionary Status}
\begin{figure}[h!]
\begin{center}
\includegraphics[width=2.3in]{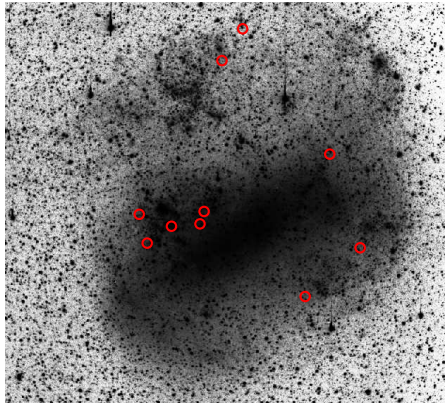} 
\caption{The locations of the WN3/O3s within the LMC.}
     \label{fig:locations}
\end{center}
\end{figure}

Now that we have a good handle on their physical parameters, we are currently investigating possible WN3/O3 progenitors as well as their later stages of evolution. Figure~\ref{fig:locations} shows where the WN3/O3s are located within the disk of the LMC. If they were all grouped together, we might assume that they formed out of the same stellar nursery. Instead, they are pretty evenly spaced across the LMC.

There are other non-binary WRs with hydrogen absorption lines denoted as WNha stars. \cite[Martins et al. (2013)]{Martins13} has argued that the majority of these stars evolved through quasi-homogeneous evolution. In this case, the stars have high enough rotational velocities that the material in the core mixes with the material in the outer layers. This creates CNO abundances near equilibrium, much as we find in the WN3/O3s. However, the WN3/O3s have relatively low rotational velocities ($V_{\rm{rot}} \sim 150$ km s$^{-1}$) and extremely low mass-loss rates. It is difficult (if not impossible) to imagine an scenario where a star could begin its life with a large enough rotational velocity (typically $\sim 250$ km s$^{-1}$) to produce homogeneous evolution and later slow down to 150 km s$^{-1}$ given the small mass-loss rates \cite[(Song et al. 2016)]{homogen}. Instead, a homogeneous star must have either a larger rotational velocity or a larger mass-loss rate. So, at this point we can rule out quasi-homogeneous evolution.

\section{Next Steps}
\cite[Drout et al.\ (2016)]{Drout16} looked at various types of supernovae progenitors and found that WN3/O3s, with their low mass-loss rates and high wind velocities, might be the previously-unidentified progenitors of Type Ic-BL supernovae. A subset of these Type Ic-BL supernovae then turn into long-duration gamma ray bursts which are preferentially found in low metallicity environments \cite[(Vink et al. 2011)]{LDGRB}. Thus, we are currently in the process of investigating any metallicity dependence for the formation of WN3/O3s. 

\begin{figure}[h!]
\begin{center}
\includegraphics[width=2.3in]{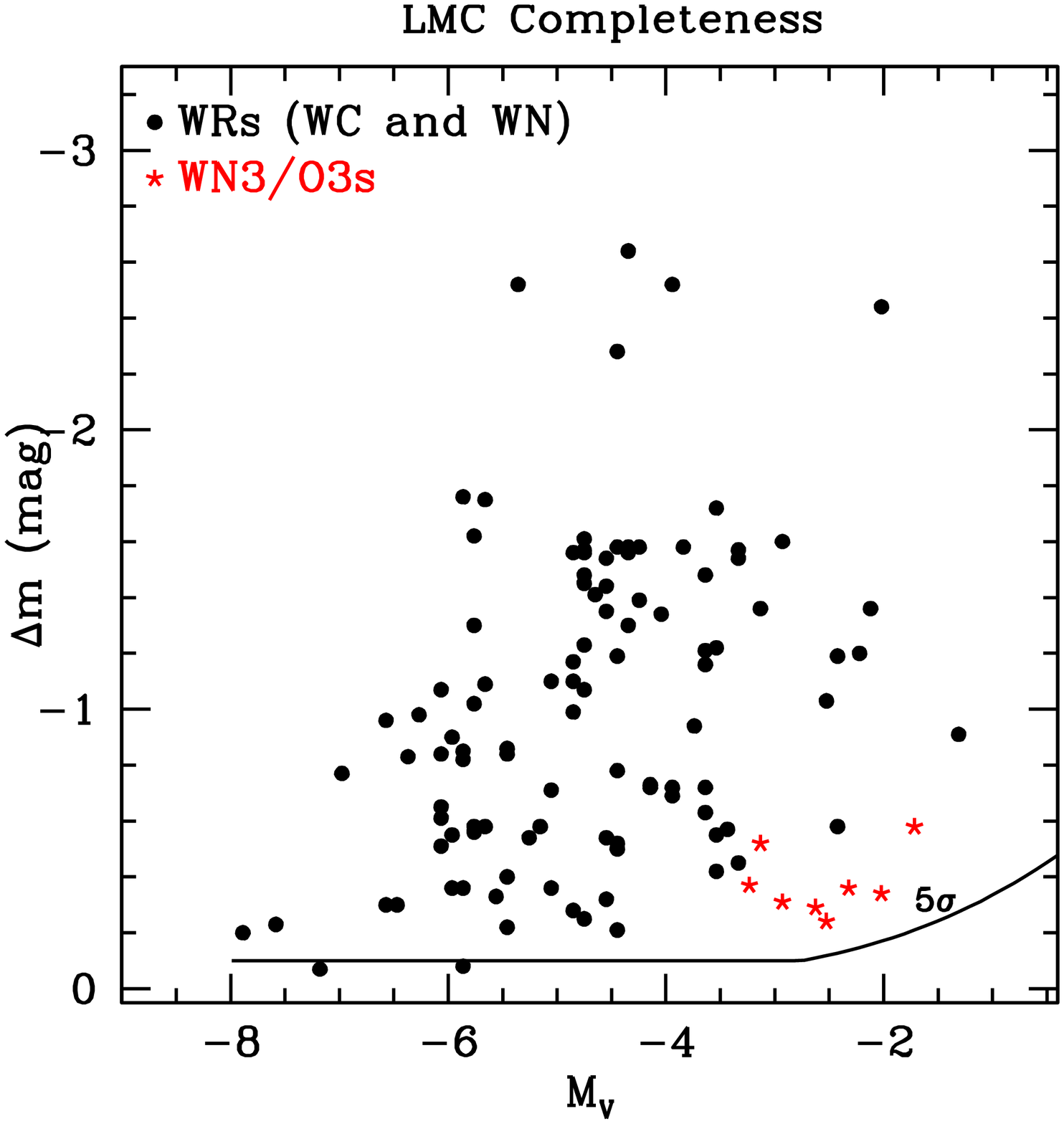} 
\includegraphics[width=2.3in]{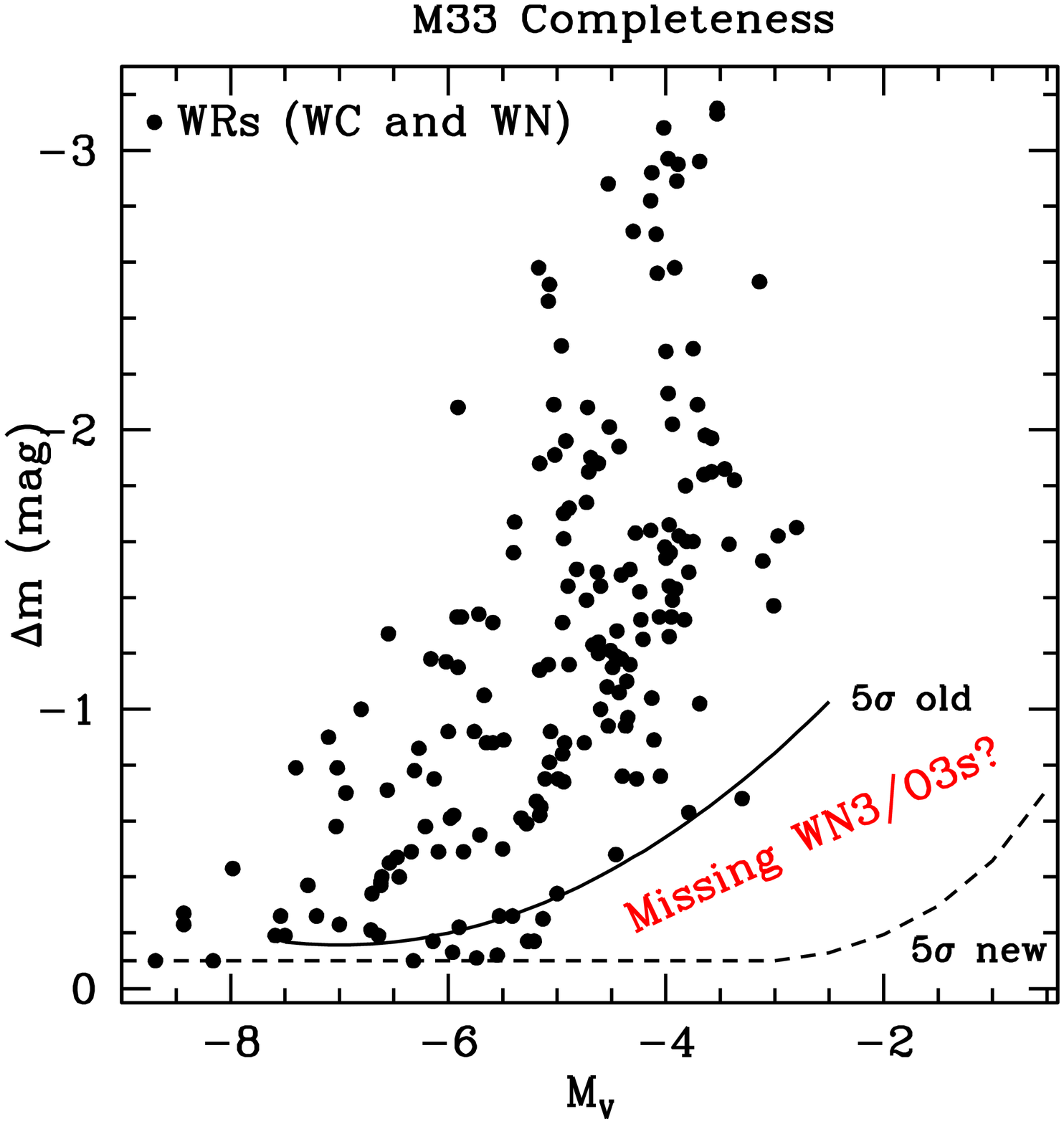} 
\caption{Magnitude difference (emission minus continuum filter) vs.\ absolute magnitude showing that our previous M33 surveys did not go deep enough to detect WN3/O3s.}
     \label{fig:M33}
\end{center}
\end{figure}

So far we have only found them in the low metallicity LMC. Given the high number of WRs currently known in the Milky Way, we would expect to have found at least a few of them. However, these dim WRs with strong absorption lines have not been found elsewhere. To investigate any metallicity dependence, we have decided to begin another search for WRs in M33 which has a strong metallicity gradient. We previously conducted a search for WRs in M33 but our previous survey simply did not go deep enough, as Figure~\ref{fig:M33} shows. So, there may be an entire yet undiscovered population of WN3/O3s within M33. 

We continue to discuss the evolutionary status of these WN3/O3s with our theoretician colleagues while also observationally attempting to determine where these stars come from and what they turn into. By searching for them in M33 we should be able to constrain their metallicity dependence and gain further insight into their evolution.

\end{document}